\documentclass[useAMS, usenatbib]{mn2e}
\usepackage{times}
\usepackage{graphicx}
\usepackage{epsfig}
\usepackage[T1]{fontenc}
\usepackage{ifpdf}
\usepackage[varg]{txfonts}
\usepackage{comment}
\usepackage{multirow}
\usepackage{multicol}
\usepackage{xcolor}

\newcommand{\Msun}      {\mbox{\,$M_{\mathord\odot}$}}

\title[Classifying IGR~J18007--4146 with XMM and NuSTAR]{Classifying IGR~J18007--4146 as an intermediate polar using {\em XMM} and {\em NuSTAR}}

\author[Coughenour et al.]{
Benjamin M. Coughenour$^{1}$\thanks{E-mail: coughenour@berkeley.edu},
John A. Tomsick$^{1}$,
Aarran W. Shaw$^{2}$,
Koji Mukai$^{3, 4}$, \newauthor
Ma\"{i}ca Clavel$^{5}$,
Jeremy Hare$^{6}$\thanks{NASA Postdoctoral Program Fellow},
Roman Krivonos$^{7}$,
and Francesca M. Fornasini$^{8}$
\\ \\
$^{1}$Space Sciences Laboratory, 7 Gauss Way, University of California, Berkeley, CA 94720-7450, USA \\
$^{2}$Department of Physics, University of Nevada, Reno, NV 89557, USA \\
$^{3}$CRESST and X-ray Astrophysics Laboratory, NASA Goddard Space Flight Center, Greenbelt, MD 20771, USA \\
$^{4}$Department of Physics, University of Maryland, Baltimore County, 1000 Hilltop Circle, Baltimore, MD 21250, USA \\
$^{5}$Universit\'{e} Grenoble Alpes, CNRS, IPAG, F38000 Grenoble, France \\
$^{6}$NASA Goddard Space Flight Center, Greenbelt, MD 20771, USA \\
$^{7}$Space Research Institute, Russian Academy of Sciences, Profsoyuznaya 84/32, 117997 Moscow, Russia \\
$^{8}$Stonehill College, 320 Washington Street, Easton, MA 02357, USA }

\begin{document}

%\date{Accepted 1988 December 15. Received 1988 December 14; in original form 1988 October 11}

\def\lsim{\mathrel{\lower .85ex\hbox{\rlap{$\sim$}\raise
.95ex\hbox{$<$} }}}
\def\gsim{\mathrel{\lower .80ex\hbox{\rlap{$\sim$}\raise
.90ex\hbox{$>$} }}}

%\pagerange{\pageref{firstpage}--\pageref{lastpage}} \pubyear{2014}

\maketitle

\label{firstpage}

\begin{abstract}

\noindent
Many new and unidentified Galactic sources have recently been revealed by ongoing hard X-ray surveys. A significant fraction of these have been shown to be the type of accreting white dwarfs known as cataclysmic variables (CVs). Follow-up observations are often required to categorize and classify these sources, and may also identify potentially unique or interesting cases. One such case is IGR~J18007--4146, which is likely a CV based on follow-up {\em Chandra} observations and constraints from optical/IR catalogs. Utilizing simultaneous {\em XMM-Newton} and {\em NuSTAR} observations, as well as the available optical/IR data, we confirm the nature of IGR~J18007--4146 as an intermediate polar type CV. Timing analysis of the {\em XMM} data reveals a periodic signal at $424.4\pm0.7$\,s that we interpret as the spin period of the white dwarf. Modeling the 0.3--78\,keV spectrum, we use a thermal bremsstrahlung continuum but require intrinsic absorption as well as a soft component and strong Fe lines between 6 and 7\,keV. We model the soft component using a single-temperature blackbody with $kT = 73^{+8}_{-6}$\,eV. From the X-ray spectrum, we are able to measure the mass of the white dwarf to be $1.06^{+0.19}_{-0.10}$\,\Msun, which means IGR~J18007--4146 is more massive than the average for magnetic CVs.

\end{abstract}


\section{Introduction}

Since its launch in 2002, the {\em International Gamma-Ray Astrophysics Laboratory (INTEGRAL)} has covered the entire sky with increasing depth above 20\,keV, detecting over 1000 distinct hard X-ray sources. While most of these sources were previously known, having been detected with soft X-ray instruments, {\em INTEGRAL's} wide field-of-view and broad high-energy coverage has led to the discovery of hundreds of new hard X-ray sources, referred to as {\em INTEGRAL} Gamma-ray (IGR) sources. While most of these are AGN, many are Galactic accreting sources such as cataclysmic variables (CVs) or neutron star and black hole X-ray binaries. Still, many sources ($>$\,200) remain unclassified \citep{bird16, krivonos17, krivonos21}.

A clear strategy to classify unknown sources is to use follow-up observations, particularly in the soft X-ray band where these sources may be localized to better accuracy than {\em INTEGRAL's} $\sim$arcminute angular resolution. In order to study and classify Galactic IGR sources, \cite{tomsick20} used {\em Chandra} follow-up observations of unclassified IGR sources from the \cite{bird16} catalog within $15^{\circ}$ of the Galactic plane. Of the 15 selected unclassified targets, 10 likely (in some cases certain) {\em Chandra} counterparts were detected, among which IGR~J18007--4146, IGR~J15038--6021, and IGR~J17508--3219 are most likely Galactic sources based on {\em Gaia} parallax distances. Of these, the former 2 are likely CVs or intermediate polars (IPs), with the possibility of high-mass donor stars being ruled out by their optical/IR counterparts \citep{tomsick20}. As was the case with other CVs discovered in the high-energy band of {\em INTEGRAL}, such as the high-mass magnetic CV IGR~J14091--6108 \citep{tomsick16}, these candidate CVs are interesting sources worthy of additional study. In this paper, we focus our investigation on the source IGR~J18007--4146.

In magnetic CVs (mCVs), the accretion disc is truncated well above the white dwarf (WD) surface, if one forms at all (truncated discs are typical of IPs but not of polars). These systems produce hard X-rays due to accretion along the magnetic field lines of the WD. When this accretion column slows from its free-fall velocity very near the WD surface, the resultant shock heats the material to extraordinary temperatures, and as the accreted material cools, it does so via bremsstrahlung radiation, producing hard X-rays \citep[see][for a review of CVs]{patterson94, warner03}. In particular, when these sources are observed in the hard X-ray band with instruments like {\em NuSTAR}, measurements of the shock temperature (a cutoff in the high-energy spectrum) provide a direct estimate of the WD mass \citep{suleimanov16, suleimanov18, hailey16, shaw18, shaw20}. Since a higher bremsstrahlung temperature is a natural consequence of a more massive WD, IPs detected in the {\em INTEGRAL} band may be biased towards higher mass WDs.

To confirm the CV nature of IGR~J18007--4146 (hereafter IGR~J18007) and closely investigate is properties, we carried out joint {\em XMM-Newton} and {\em NuSTAR} observations of the source and investigated its spectral and timing properties. In Section~2, we describe these observations and the data reduction. In Section~3, we describe our timing analysis and search for periodicity in the source light curve (3.1) as well as our modeling of the source energy spectra (3.2). Finally, we discuss these results and their implications in Section~4.

%%%%% TABLE 1 %%%%
\begin{table*}
%\centering
\caption{Observations of IGR~J18007--4146 \label{tab:obs}}
\begin{minipage}{\linewidth}
\begin{center}
\begin{tabular}{cccccccc} \hline \hline
Observatory  & ObsID         & Instrument & Start Time (UT) & End Time (UT) & Exposure Time (ks)\\ \hline\hline
{\em XMM}    & 0870790201	& pn 		& 2020 Oct 3, 15.24 h & 2020 Oct 3, 23.20 h & 23.2\\
''           & ''            			& MOS1 	& 2020 Oct 3, 15.79 h & 2020 Oct 3, 23.29 h & 25.1\\
''           & ''            			& MOS2 	& 2020 Oct 3, 16.20 h & 2020 Oct 3, 23.29 h & 23.7\\
{\em NuSTAR} & 30601017002 & FPMA 	& 2020 Oct 3, 9.44 h & 2020 Oct 4, 6.02 h & 39.9\\
''           & ''            & FPMB       & ''                    & ''                    & 39.6\\
\hline
\end{tabular}
\end{center}
\end{minipage}
\end{table*}

\section{Observations and Data Reduction}

Observations of IGR~J18007 were carried out simultaneously with the {\em X-ray Multi-Mirror Mission (XMM-Newton)} and the {\em Nuclear Spectroscopic Telescope Array (NuSTAR)} beginning on 2020 October 3, with the {\em NuSTAR} observation lasting until 2020 October 4. In the case of {\em XMM-Newton} \citep{jansen01}, the duration of the observation was 37.5\,ks. The longer {\em NuSTAR} \citep{harrison13} observation produced $\sim$40\,ks of exposure in each Focal Plane Module (known as FPMA and FPMB) after dead-time corrections and Earth occultations. These observation details are summarized in Table~\ref{tab:obs}.

%%%%% FIGURE 1 %%%%%
\begin{figure}
\begin{center}
    \includegraphics[width=8cm,trim=1cm 0.3cm 0.2cm 0.7cm, clip]{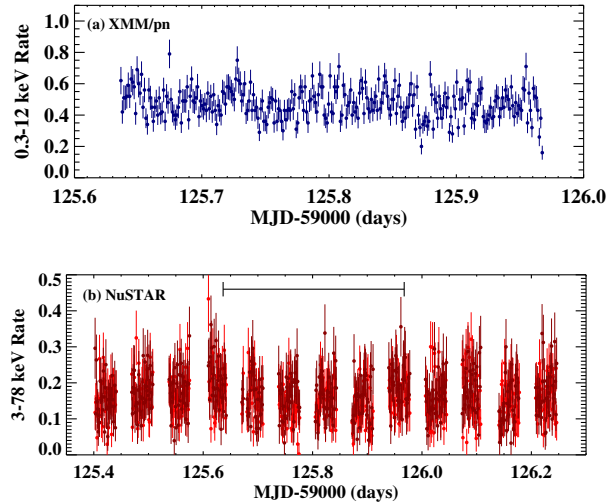}
    \caption{\small Light curves shown for (a) {\em XMM} and (b) {\em NuSTAR}, with FPMA/B shown in light/dark red, respectively. Note the different x-axis scale between the two light curves; the shorter duration of the {\em XMM} observation is illustrated at the top of the {\em NuSTAR} light curve with a horizontal bar.
\label{fig:lc}}
\end{center}
\end{figure}

\subsection{{\em XMM}}

We reduced the EPIC/pn \citep{struder01} and EPIC/MOS \citep{turner01} data with the XMM Science Analysis Software v18.0.0 using the standard analysis procedures provided on-line\footnote{See https://www.cosmos.esa.int/web/xmm-newton/sas-threads}. All three detectors (pn, MOS1, and MOS2) were operated in Full Window mode with the medium optical-blocking filter. We used the SAS routine {\ttfamily epatplot} to determine that no significant photon pile-up is present in pn, MOS1, or MOS2. For the pn, we filtered the photon event lists requiring the FLAG parameter to be zero and the PATTERN parameter to be less than or equal to 4. For MOS, we used the filtering expression, "\#XMMEA\_EM \&\& (PATTERN<=12)."

The pn and MOS images show a single bright source at the {\em Chandra} position of IGR~J18007. We extracted a 0.3--12\,keV pn light curve using a circular source region centred on IGR~J18007 with a radius of 640 pixels ($32^{\prime\prime}$) along with a background light curve. We specified a time resolution of 100\,s, and the background-subtracted pn light curve is shown in the top panel of Figure~\ref{fig:lc}. Although the source does not show any dramatic variability, a low-level of variability is apparent.

We also extracted a 10--12\,keV pn light curve with 100\,s time bins including photons from the entire field-of-view. This is done to determine if there are background particle flares, and we find that this observation includes a few flares. We created a good time interval (gti) filter to include only the times when the count rate in the 10--12\,keV pn light curve is less than 0.5\,c/s. After filtering out the background flares, we extracted source and background energy spectra. We used the same gti filter for pn, MOS1, and MOS2 in order to maintain near-simultaneity between the instruments. After using {\ttfamily evselect} to extract source and background spectra, we used {\ttfamily backscale} to set the background scaling parameter, which accounts for the size of the extraction regions and any bad pixels in the regions. In addition, we used {\ttfamily rmfgen} and {\ttfamily arfgen} to produce the response matrices. We binned the spectra to require a signal-to-noise ratio of 5 in each bin (except the highest-energy bin). The fitting of the spectra is described in Section 3.2.

\subsection{{\em NuSTAR}}

{\em NuSTAR} ObsID 30601017002 was reduced using NUSTARDAS v2.0.0 and CALDB 20210427. Standard data products were produced with \texttt{nupipeline}, including 3--78\,keV images for both FPMA and FPMB. For the timing analysis, barycenter corrected event files were created using \texttt{nuproducts}. Using these images, circular regions of 60\,arcsec and 90\,arcsec in radius were chosen for the source and background regions. IGR~J18007 was clearly visible, and the background region was positioned to be on the same detector chip as the source. The 3--78\,keV spectra were then extracted using \texttt{nuproducts} for both FPMA and FPMB, and these spectra were then each binned using \texttt{grppha} to require a signal-to-noise ratio of 5 in each bin (except the highest-energy bin), as was done for the 0.3--12\,keV {\em XMM-Newton} spectra (see Section 2.1). The background-subtracted light curves for {\em NuSTAR} FPMA and FPMB are shown together in the bottom panel of Figure~\ref{fig:lc}.

\section{Results}

\subsection{X-ray Timing}

%%%%% FIGURE 2 %%%%%
\begin{figure*}
\begin{center}
    \includegraphics[width=12cm, trim=0.8cm 0.25cm 0.5cm 0cm, clip]{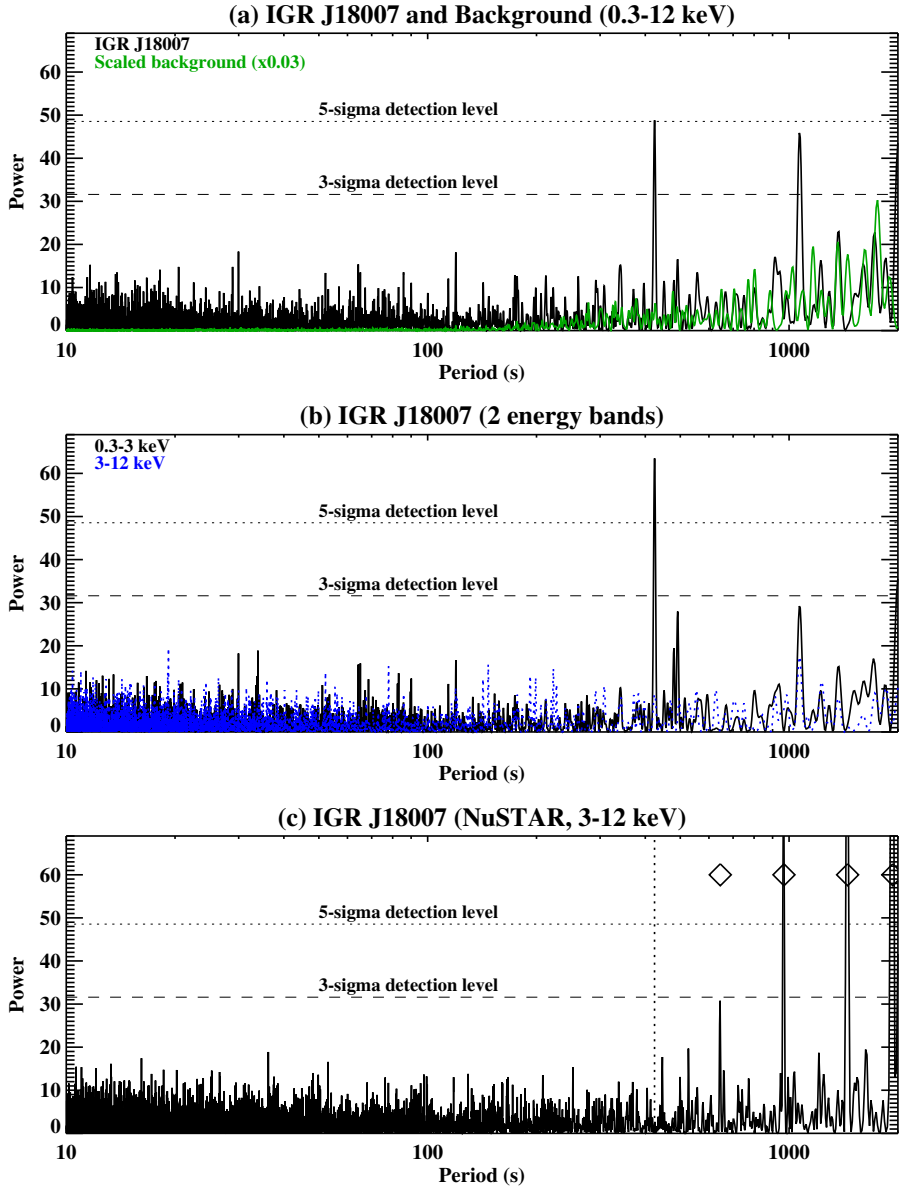}
    \caption{\small {\em (a)} The results of a $Z^{2}_{1}$ (Rayleigh) test periodicity search using all 0.3--12\,keV photons in the circular source (J18007) regions for the pn, MOS1, and MOS2 instruments onboard {\em XMM}. The green curve shows the scaled background periodogram. {\em (b)} Periodograms for J18007 in the 0.3--3\,keV and 3--12\,keV energy bands, showing the high significance of the $424.4\pm 0.7$\,s period in the soft band. {\em (c)} Periodogram for the 3--12\,keV summed FPMA/B light curve from {\em NuSTAR}, which does not detect the 424.4\,s period signal seen with {\em XMM} (vertical dotted line). Open diamonds denote harmonics of the {\em NuSTAR} orbital period.
\label{fig:period}}
\end{center}
\end{figure*}

For the {\em XMM}/EPIC instruments (pn, MOS1, and MOS2), we made event lists including the 0.3--12\,keV photons in the same circular source regions used for the spectral analysis. We corrected the timestamp for each event to the time at the solar system barycenter using the SAS routine {\ttfamily barycen}. Given that the results of \cite{tomsick20} strongly suggest that IGR~J18007 is an IP (this is also supported by our spectral analysis, see Section~3.2), we carried out a search for a periodic signal, which is often seen at the WD spin period for IPs. We used the $Z^{2}_{1}$ test \citep{buccheri83} to search for a periodic signal in 20,000 evenly-spaced frequency bins between $5\times 10^{-4}$ and 0.1\,Hz, corresponding to periods between 10 and 2000\,s. The range of periods searched is based on typical white dwarf spin periods in IPs \citep{deMartino20}, and the number of bins is set to oversample the number of independent frequencies by a factor of eight. Although the specific oversampling number is not critical, some oversampling is necessary to ensure that the peaks seen are not biased by the exact value of the starting frequency.

The results of the periodicity search are shown in Figure~\ref{fig:period}a. The false alarm probability (FAP) for this test is given by $N_{\rm trials}$\,$e^{-S/2}$ where $S$ is the peak signal power and $N_{\rm trials}$ is the number of frequencies searched. In this case, the 3$\sigma$ detection level, corresponding to FAP $= 2.7\times 10^{-3}$, is at $S=31.6$. There are two peaks in the periodogram above the 3$\sigma$ level. The strongest (at $S=48.8$) is at $424.4\pm 0.7$\,s, and a slightly weaker ($S=46.2$) signal is at $1069.5\pm 3.0$\,s. The 1$\sigma$ uncertainties on the periods are found by determining where the signal drops to a signal power of $S$--1. The FAPs for the 424.4 and 1069.5\,s signals are $5.1\times 10^{-7}$ (5$\sigma$) and $1.9\times 10^{-6}$ (4.75$\sigma$), respectively.

We also performed a periodicity search with the {\em NuSTAR} data using the $Z^{2}_{1}$ test and the same 20,000 trial periods used for {\em XMM}, which, as for {\em XMM}, puts the 3-$\sigma$ significance threshold at $S=31.6$. We used event lists that include photons detected by FPMA and FPMB in three energy bands (3--6, 3--12, and 3--78\,keV). All the peaks in the periodogram that exceed $S=31.6$ are harmonics of the {\em NuSTAR} orbital period. Peaks are seen at 968\,s, 1452\,s, and 1936\,s, and they are expected due to Earth occultations. There are no significant peaks in the {\em NuSTAR} periodograms at either of the {\em XMM} periods. At 424.4 and 1069.5\,s, the 3--12\,keV {\em NuSTAR} periodogram has values of $S = 1.9$ and $S = 3.7$, respectively, and this periodogram is given in Figure~\ref{fig:period}c.

Given the lack of detection of the {\em XMM} periods in the {\em NuSTAR} periodograms, we made additional {\em XMM} periodograms in the 0.3--3 and 3--12\,keV energy bands to check on the energy dependence of the signals. Figure~\ref{fig:period}b shows that the 424.4\,s signal has $S=63.5$ at 0.3--3\,keV, which, even after accounting for the extra trials, corresponds to a significance above 6$\sigma$. In comparison, the 3--12\,keV periodogram has a value of $S=8.8$ at 424.4\,s. The fact that the 424.4\,s signal is not detected in the 3--12\,keV band but is strong at 0.3--3\,keV indicates that it is has an extremely soft spectrum, which is consistent with the fact that it is not seen by {\em NuSTAR}. The 1069.5\,s signal is not detected at the 3$\sigma$ level in either energy band individually, but the ratio of the strength of the 0.3--3\,keV peak to the 3--12\,keV peak is much smaller than for 424.4\,s, indicating that if it is a real signal, it has a much harder spectrum. While the softness of the 424.4\,s signal provides a plausible explanation for why we do not see it in the {\em NuSTAR} periodogram, it is somewhat surprising that we do not see any evidence for a peak at 1069.5\,s with {\em NuSTAR}. Thus, we focus on the 424.4\,s signal in the rest of this section.

The 424.4\,s signal is seen in each instrument individually, increasing our confidence that it is a real signal from IGR~J18007. However, we carried out an additional check because there is variability in the background, and the $Z^{2}_{1}$ test uses a photon list which is not background subtracted. In addition, we did not filter the photon list for times of high background because putting gaps in the data can also cause spurious signals. Thus, we carried out an identical analysis to the $Z^{2}_{1}$ test that led to the detection of the 424.4\,s signal in IGR~J18007, but we used a photon list from the next brightest source in the field. The comparison source is much fainter than IGR~J18007, making it a good test of whether the background can induce a spurious signal. The result for the comparison source is that the power of the strongest peak is $S=16.8$, which is consistent with a non-detection. Furthermore, at 424.4\,s, the power is $S=2.4$, confirming that the background variability does not induce a peak at the IGR~J18007 detection period. Another demonstration that the background is not inducing the 424.4\,s is that we made a background periodogram taken from large regions of the pn, MOS1, and MOS2 fields, and this is shown in Figure~\ref{fig:period}a.

\begin{figure}
\begin{center}
    \includegraphics[width=8.5cm, trim=1cm 0.25cm 0.2cm 0.5cm, clip]{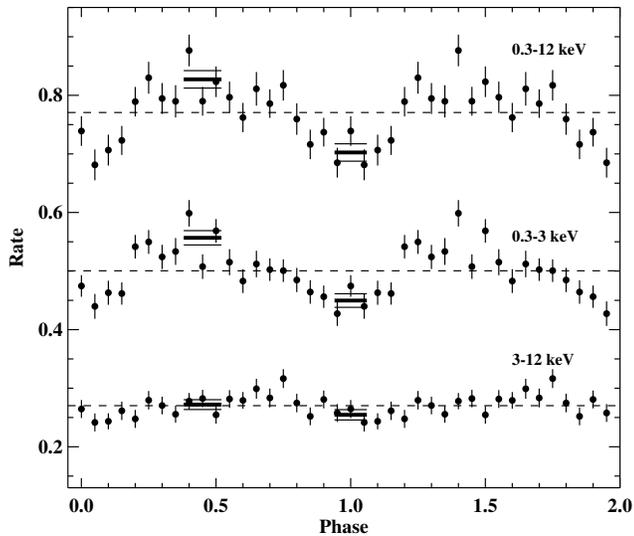}
    \caption{    \small {\em XMM} light curves folded on a period of 424.4\,s. The count rate plotted is the sum of the pn, MOS1, and MOS2 rates, in three different energy bands. On top is the full count rate (0.3--12\,keV), which is the sum of the soft (0.3--3\,keV, middle) and hard (3--12\,keV, bottom) count rates. The thick solid horizontal lines show the average of the three points at the minimum and maximum phases, as defined by the phase of the full count rate. The thin solid horizontal lines show the 1$\sigma$ uncertainties on the averages. Two cycles of the period are shown for clarity.
\label{fig:folded_LC}}
\end{center}
\end{figure}

\begin{figure}
\begin{center}
    \includegraphics[width=6.5cm, trim=1cm 0.4cm 0.5cm 0.8cm, clip]{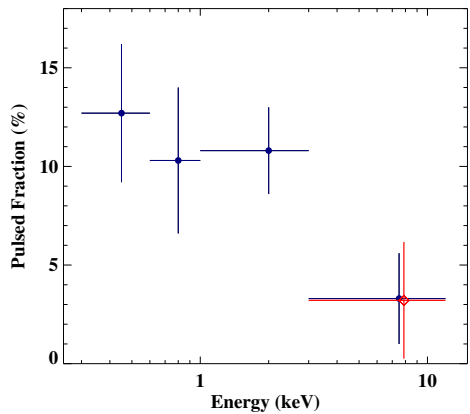}
    \caption{\small The pulsed fraction vs. energy for {\em XMM} (filled circles) and {\em NuSTAR} (red diamond). The {\em XMM} points include the pn, MOS1, and MOS2 data, while {\em NuSTAR} shows the combined FPMA/B. The pulsed fraction is determined from the folded light curves, and is defined as maximum minus minimum divided by maximum plus minimum. The uncertainties shown are at the 1$\sigma$ confidence level.\label{fig:periodSpec}}
\end{center}
\end{figure}

We carried out an epoch folding analysis with the 0.3--12\,keV {\em XMM} pn and MOS light curves. This provides another check on the detection of the period since we perform the folding with a background-subtracted light curve. Folding at a range of trial periods yields peaks for pn and MOS consistent with the 424.4\,s signal. We combined the pn, MOS1, and MOS2 light curves after trimming them to have the same start and stop times (which we also did for the $Z^{2}_{1}$ test described above). Figure~\ref{fig:folded_LC} shows the result of folding specifically on the 424.4\,s period, and a significant modulation can be seen in the 0.3--3\,keV and 0.3--12\,keV energy bands, but less so in the 3--12\,keV band. The maxima and minima and their uncertainties are marked, and we calculate the pulsed fraction as maximum minus minimum divided by maximum plus minimum. We find a 0.3--12\,keV pulsed fraction of 8.2\%$\pm 1.4$\%. We repeated the analysis in four energy bands using the maximum and minimum phase ranges from the 0.3--12\,keV analysis, and Figure~\ref{fig:periodSpec} summarizes these results. There is no evidence for a difference in amplitude between 0.3 and 3\,keV with the average being 11\%, but the 3--12\,keV amplitude is significantly lower with a value of 3.2\%$\pm 2.3$\%, which is consistent with a non-detection.

We also performed a folding analysis of the {\em NuSTAR} light curves specifically at the 424.4\,s period detected in the {\em XMM} data. We first shifted the {\em NuSTAR} light curve by matching the start time to the phase zero epoch found with {\em XMM}. Then, we calculated the maximum and minimum values in the folded light curve using the maximum and minimum phase ranges from the 0.3--12\,keV {\em XMM} analysis to determine the pulsed fraction for the 3--12\,keV {\em NuSTAR} band. This analysis was performed separately for FPMA and FPMB. Although no statistically significant detection was obtained, the measurements are consistent with {\em XMM}. In the 3--12\,keV band, we find 3.2\%$\pm 3.0$\%, and this point is shown along with the {\em XMM} values in Figure~\ref{fig:periodSpec}.

%%%%% FIGURE 3 %%%%%

%%%%% FIGURE 4 %%%%%

\subsection{X-ray Spectrum}

For spectral analysis, we used the XSPEC package v12.11.1 \citep{arnaud96} and performed the fitting with $\chi^2$ minimization. Spectra from each of the 5 separate detectors, EPIC/pn and EPIC/MOS[1/2] from {\em XMM-Newton}, as well as FPM[A/B] from {\em NuSTAR}, were fit together using the XSPEC model \textsc{constant}, with the factor for the EPIC/pn spectrum set fixed to 1.0 and all others allowed to vary. This is in order to account for any calibration differences between the different instruments. Interstellar absorption was taken into account using the model \textsc{tbabs}, with \cite{wilms01} abundances. Errors quoted in our spectral analysis represent the 90\% confidence level.

Initial fits using an absorbed power-law with a high-energy cutoff (\textsc{cutoffpl} in XSPEC) showed that the source spectrum is very hard up to $\sim$\,30\,keV, with a photon index of $0.93\pm0.03$. This value is consistent with the reported results of \cite{tomsick20} using {\em Chandra} and {\em INTEGRAL} data, and suggests that IGR~J18007 is either a CV or HMXB. Residuals in the 6--7\,keV range and below 1\,keV imply the presence of Fe line emission and a soft thermal component. Indeed, including a low-temperature ($\sim$\,0.1\,keV) blackbody (with \textsc{bbodyrad}) as well as a broad Gaussian Fe line at 6.5\,keV dramatically improved the fit to $\chi^2 = 929$ with 845 degrees of freedom (dof).

Considering that IGR~J18007 is likely an IP, we continue our spectral analysis using a more physical model. Bremsstrahlung emission is expected to produce the majority of the continuum source emission, as is generally the case for magnetic CVs \citep{mukai17}, and we chose the model \textsc{bremss} to represent the broad continuum, modified by reflection off of neutral material (in this case, the WD surface) using \textsc{reflect} in XSPEC \citep{mz95}. We replaced the single, broad emission line measured at $6.5\pm0.4$\,keV from our initial fits with 3 individual Gaussians fixed at 6.40, 6.70, and 6.97\,keV to represent the reflected neutral Fe~K$\alpha$ line alongside the Fe\,\textsc{xxv} and Fe\,\textsc{xxvi} ionization states, respectively. At first we tied the Gaussian $\sigma$ values together, to prevent a single line from broadening to cover the others, but found their widths to be poorly constrained and unrealistically high (with $\sigma \sim 200$\,eV). To fix this we froze the width of each line to be $\sigma\,=\, 50$\,eV to better represent the expected narrow emission lines, and we found all 3 lines to be significantly detected when using the narrow Gaussians. Each line was inconsistent with a zero normalization at $>4\sigma$ as calculated using \texttt{steppar} in XSPEC.

At first, this more physical model provided a poor fit to the data, with the model deviating significantly from the spectra below $\sim$\,3\,keV. An additional feature in the residuals is a noticeable bump at $\sim$\,16\,keV, though this has no clear physical interpretation, and we found that the presence of an additional Gaussian at 16\,keV is not highly significant. As for the low-energy residuals, we included intrinsic absorption using a partial-covering absorber with the model \textsc{pcfabs}. This immediately made a dramatic improvement to the fit, with a reduction in the fit statistic $\Delta\chi^2 = 144$ for 2 fewer dof. Some intrinsic partial covering absorption is expected in mCVs, because the cooling post-shock accretion material near the WD surface is beneath the rest of the accelerating accretion flow, and may be covered or absorbed depending on the accretion geometry and line of sight \citep{norton89}. The additional absorption model component is therefore both physically and statistically motivated, and provides more evidence supporting the IP nature of IGR~J18007. We also tested whether the partial covering absorption is able to fit the low-energy spectra without the need for a soft component, but we found that both are required to produce even a reasonable fit to the spectra. This reflected bremsstrahlung continuum model, with partial covering absorption, 3 distinct Fe lines, and a soft component, provided an excellent fit to the data, with $\chi^2/{\rm dof} = 840/842$, and we measure a bremsstrahlung plasma temperature of $24.2^{+7.3}_{-2.0}$\,keV.

As a final test of the soft component, we also fit the spectra using the cooling flow model \textsc{mkcflow} \citep{mushotzky88}, which includes atomic emission lines as well as an optically thin thermal continuum. Since \textsc{mkcflow} reproduces the 6.70 and 6.97\,keV Fe lines, we removed the two higher energy Gaussians but kept the 6.40\,keV line produced by reflection. A series of emission lines below 1\,keV could conceivably reproduce the soft excess without the need for a blackbody, and so by modeling the spectra with \textsc{mkcflow} we tested the physicality of the soft blackbody component. We found that the blackbody still significantly improves our fits, and is required to satisfactorily fit the source spectrum alongside the partial covering absorption. Furthermore, we found that the blackbody temperature when using the cooling flow model does not significantly change from our results with a bremsstrahlung continuum, with both measurements consistent with $kT = 71$\,eV at the 90\% confidence level. The \textsc{mkcflow} model also gives a measure of the maximum plasma temperature, and we find $T_{max}=67^{+13}_{-20}$\,keV, which is consistent with the hard-coded upper limit of the model at 79.9\,keV.

%%% TABLE 2 %%%
\begin{table}
\caption{Spectral Results for PSR Model\label{tab:params}}
\begin{minipage}{0.9\columnwidth}
\begin{tabular}{lccc} 
\multicolumn{4}{c}{\textsc{constant$^\ast$tbabs$^\ast$pcfabs$^\ast$(reflect$^\ast$ipolar + bbodyrad + 3\,gauss)}} \\
\hline \hline
Model & Parameter & Units & Value \\ \hline
\textsc{tbabs} & $N_{\rm H}$ & $10^{22}$\,cm$^{-2}$ & $0.03\pm0.02$ \\
\textsc{pcfabs} & $N_{\rm H}$ & $10^{22}$\,cm$^{-2}$ & $11.7\pm2.3$ \\
\textsc{pcfabs} & fraction & --- & $0.63^{+0.05}_{-0.12}$ \\ \hline
  & $\Omega/2\pi$ & --- & 1.0 \footnote{Fixed.} \\
\textsc{reflect} & $A$ & --- & $0.5^{+0.7}_{-0.4}$
\footnote{Abundance of elements heavier than He, including Fe, relative to solar values.} \\
%$A_{\rm Fe}$\footnote{The abundance of Fe relative to solar. Set equal to $A$.} & --- & $1.64^{+1.51}_{-0.99}$\\
  & $\cos{i}$ & --- & $0.65^{+0.30}_{-0.41}$ 
  \footnote{Model upper limit of 0.95 within 90$\%$ confidence interval.} \\ \hline
  & $M_{\rm WD}$ & \Msun & $1.06^{+0.19}_{-0.10}$ \\
\textsc{ipolar} & fall height & $R_{\rm WD}$ & 17.4 
\footnote{Fixed to vary with WD mass and spin period} \\
  & $N_{\rm PSR}$ & $10^{-29}$ & $8.8^{+6.6}_{-6.5}$ \\ \hline
\multirow{2}{*}{\textsc{bbodyrad}}
  & $kT$ & eV & $73^{+8}_{-6}$ \\
  & $N$ & $10^3$ & $3.9^{+4.9}_{-2.6}$ \\ \hline
  & $E_1$ & keV & 6.4~$^{a}$ \\
\textsc{gauss}$_1$  & $\sigma_1$ & keV & 0.05~$^{a}$ \\
  & $N_1$ & $10^{-6}\;$ph\,cm$^{-2}$\,s$^{-1}$ & $11.9\pm1.7$ \\ \hline
  & $E_2$ & keV & 6.7~$^{a}$ \\
\textsc{gauss}$_2$  & $\sigma_2$ & keV & 0.05~$^{a}$ \\
  & $N_2$ & $10^{-6}\;$ph\,cm$^{-2}$\,s$^{-1}$ & $6.7\pm1.7$\\ \hline
  & $E_3$ & keV & 6.97~$^{a}$ \\
\textsc{gauss}$_3$ & $\sigma_3$ & keV & 0.05~$^{a}$ \\
  & $N_3$ & $10^{-6}\;$ph\,cm$^{-2}$\,s$^{-1}$ & $3.8\pm1.5$\\ \hline
  & $C_{\rm pn}$ & --- & 1.0~$^{a}$ \\
  & $C_{\rm MOS1}$ & --- & $1.00 \pm0.03$ \\
\textsc{constant}  & $C_{\rm MOS2}$ & --- & $0.99 \pm0.03$ \\
  & $C_{\rm FPMA}$ & --- & $1.06 \pm 0.04$ \\
  & $C_{\rm FPMB}$ & --- & $1.15 \pm 0.05$ \\ \hline
---  & $\chi^{2}$/dof & --- & 844.4 / 842 \\ \hline
\end{tabular}
Best-fitting results are shown for the PSR model of \cite{suleimanov16, suleimanov19}. This is included in the model description as `\textsc{ipolar}' at the top of the table. As with the bremsstrahlung model, a blackbody soft component, intrinsic absorption, and three narrow Gaussian lines are included in the model. Errors represent 90\% confidence intervals.
\vspace{-0.5cm}
\end{minipage}
\end{table}

%%%%% FIGURE 5 %%%%%
\begin{figure*}
\begin{center}
\includegraphics[width=12.5cm,trim=1.5cm 4cm 5.5cm 4.5cm, clip]{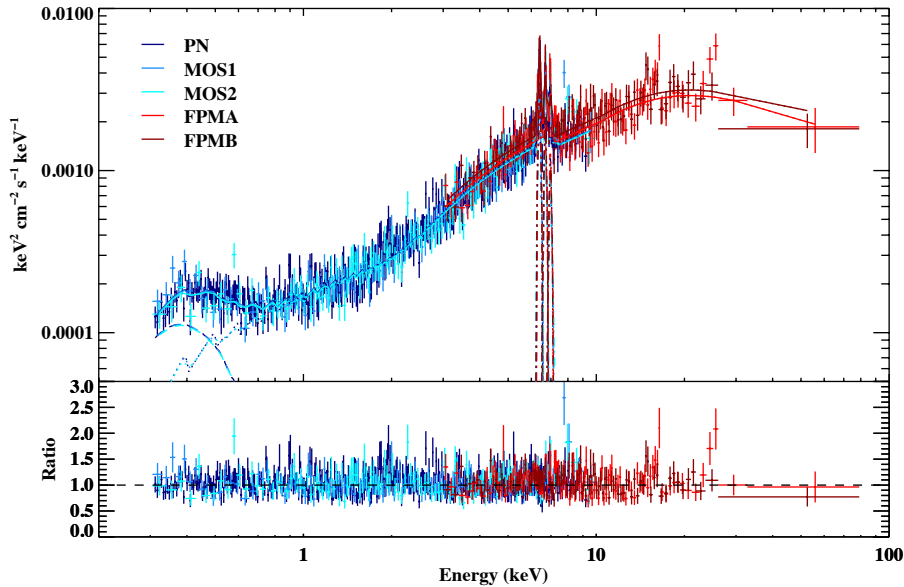}
\caption{\small Unfolded {\em XMM} and {\em NuSTAR} energy spectrum fitted with 
the PSR IP mass model (see Table~\ref{tab:params}). The model consists of a low-temperature blackbody ({\em dashed line}), the broad PSR model ({\em dotted line}), and 3 narrow Gaussians between 6 and 7\,keV ({\em dash-dotted lines}).\label{fig:spectrum}}
\end{center}
\end{figure*}

\subsubsection{Measuring the WD Mass}

In order to measure the WD mass, we replace the bremsstrahlung continuum emission with an updated version of the IP mass model of \cite{suleimanov05}, referred to as the post-shock region or `PSR' model (\textsc{ipolar} in XSPEC) and described in \cite{suleimanov16, suleimanov19}. As with our bremsstrahlung continuum model, we continue to use reflection, partial covering absorption, a blackbody soft component, and 3 Gaussian emission lines with the PSR model continuum. Previous versions of the PSR model calculated the shock temperature using the free-fall velocity of incoming material, which is a reasonable approximation for many CVs. However, by not taking into account the finite magnetospheric radius at which the accretion disc is truncated, the WD mass may be systematically underestimated \citep{suleimanov16, shaw20}. This updated PSR model takes into account a limited fall height in order to avoid such a bias. Following the method of \cite{suleimanov19} and \cite{shaw20}, we set the magnetospheric radius equal to the co-rotation radius between the WD and inner accretion disc:

\begin{equation}
R_{\rm m} \simeq R_{\rm co} = \left( \frac{ G M_{\rm WD} P^2_{\rm spin} }{ 4 \pi^2 } \right)^{1/3}
\end{equation}

The assumption that the co-rotation radius and magnetospheric radius are equal is acceptable if the IP has been persistently accreting, allowing the WD to reach spin equilibrium with material in the disc. Though we measure two periodic signals in the {\em XMM} data, we interpret the stronger and softer signal at 424.4\,s to be the WD spin period, and we set the fall height parameter of the PSR model to vary according to the WD mass and spin period, using the equation above alongside the mass-radius relationship for WDs as calculated by \cite{nauenberg72}. The only free parameters of our IP model then are the WD mass and normalization.

As was the case for the bremsstrahlung continuum model, we obtain an excellent statistical fit using the PSR model as described, with $\chi^2$/dof = 844 / 842. Our best-fitting results using the PSR IP mass model are summarized in Table~\ref{tab:params}, and a plot of the unfolded spectrum and normalized residuals is shown in Figure~\ref{fig:spectrum}. For the WD mass, we measure $1.06^{+0.19}_{-0.10}\,\Msun$, which gives a corresponding fall height of $17.4\,R_{WD} = 8.7\times10^{4}$\,km (assuming the \cite{nauenberg72} mass-radius relationship). For the soft component, we measure a blackbody temperature of $73^{+8}_{-6}$\,eV, which is consistent with our earlier results and is common for IPs and polars \citep{deMartino20}.

\section{Discussion}

IGR~J18007 was chosen for follow-up investigation as a likely CV due to its hard X-ray spectrum and faint optical counterpart \citep{tomsick20}. Our results confirm this interpretation --- with higher-quality X-ray data, we still find a hard X-ray continuum, but also a strong Fe line complex between 6 and 7\,keV, a distinct soft component (which is often seen in IPs), and intrinsic partial covering absorption. The three distinct Fe lines, with the weakest line at 6.97\,keV detected at $4\sigma$, provide evidence that the source is an accreting WD rather than a black hole or neutron star. We also detect periodic signals in the {\em XMM} data, which is expected of IPs but could also be explained by a slowly rotating neutron star in an HMXB or symbiotic X-ray binary. These possibilities are ruled out by optical/IR photometry. In \cite{tomsick20}, the {\em Chandra} position of IGR~J18007 is used to identify optical/IR counterparts in the VISTA and {\em Gaia} catalogues. Given the low extinction as measured by {\em Chandra} as well as the photometric IR colours, the authors estimated the companion temperature to be $\sim6500$\,K. This analysis is supported by the \texttt{StarHorse} database \citep{anders19}, which reports an effective temperature of 6265\,K. If the source were an HMXB or contained a late-type giant, then the companion would dominate the optical/IR flux and the  temperature derived from photometry would represent the effective temperature of the companion. With this in mind, a temperature of $\sim6500$\,K is inconsistent with an HMXB or a symbiotic X-ray pulsar, and we conclude that IGR~J18007 is indeed an IP accreting from a low-mass companion.

{\em Gaia} DR2 parallax data was instrumental in constraining the properties of the companion, but with an updated {\em Gaia} EDR3 geometric distance of $2.44^{+0.28}_{-0.26}$\,kpc \citep{gaia_edr3, bj21}, we are able to better constrain the X-ray luminosity of IGR~J18007 as well. Using this distance as well as the measured 0.3--100\,keV flux of $1.08\times10^{-11}$\,erg\,cm$^{-2}$\,s$^{-1}$ based off our PSR model, we calculate a luminosity of $7.7^{+1.9}_{-1.6}\times10^{33}$\,erg\,s$^{-1}$, which is typical of known IPs \citep{deMartino20}. In the 17--60\,keV {\em INTEGRAL} band, we measure a flux of $4.68\times10^{-12}$\,erg\,cm$^{-2}$\,s$^{-1}$, and this is comparable to the 17-year average {\em INTEGRAL} flux of $7.9\times10^{-12}$\,erg\,cm$^{-2}$\,s$^{-1}$ \citep{krivonos21}. Since CVs are naturally variable sources, this range of different flux measurements over time is to be expected.

In our timing analysis, we measure two periodic signals in the {\em XMM} data above the $3\sigma$ detection level, at $424.4\pm 0.7$\,s ($5\sigma$) and $1069.5\pm 3.0$\,s ($4.75\sigma$). Neither signal is present in the {\em NuSTAR} data, though the 424.4\,s signal is very soft and isn't detected above 3\,keV in {\em XMM}. The 1069.5\,s signal, however, is almost as significant above 3\,keV as it is in the softer X-rays (but still below $3\sigma$ in either case). Spin periods in IPs are expected to have a soft spectrum \citep{mukai17}, and a soft signal is consistent with partial covering absorption as well \citep{norton89}. We therefore adopt the softer and more significant of the two detected signals (at 424.4\,s) to be the spin period of the WD in IGR~J18007. Folding at the 424.4\,s period provides a clear pulse-profile (Figure~\ref{fig:folded_LC}), and the pulsed fraction spectrum (Figure~\ref{fig:periodSpec}) shows in detail that the spectrum is soft, and that the signal strength is consistent in both {\em XMM} and {\em NuSTAR} above 3\,keV. A spin period of 424.4\,s is well within the expected range among IPs \cite[see, e.g.,][]{deMartino20}, and the properties of the 424.4\,s signal provide one more piece of evidence that IGR~J18007 is an IP.

One curiosity of our spectral analysis worth discussing is the soft component which we model with a single-temperature blackbody. We measure a temperature of $73^{+8}_{-6}$\,eV when using the PSR model. Soft components with temperatures less than 0.1\,keV are often seen IPs \citep{mukai17, deMartino20}. Given the {\em Gaia} distance to IGR~J18007, we estimate an emitting area with a radius of $15^{+19}_{-10}$\,km, which is small with respect to the size of the WD. The physical interpretation the soft component in the spectrum of an mCV is generally a matter of debate, particularly with respect to a partial absorber. In our case, this thermal emission may come from small hotspots on the WD surface close to its magnetic poles, heated by the nearby accretion column. Although typically modeled as a single-temperature blackbody, it is unlikely these hot regions would be isothermal \citep{beuermann12}, and the interpretation of the soft component as a single-temperature blackbody is therefore physically unreliable. As an additional complication, it is unclear whether or not the soft component would then be subject to the same absorption as the hard X-rays from the accretion column, and in some instances the soft component may be an artifact of partial absorption on the continuum emission of the source. As described in Section~3.2, this is not the case for IGR~J18007, and we are confident that a soft component is present and likely represents hotspots on the WD surface.

Using the PSR model of \cite{suleimanov16}, we measure the WD mass to be $1.06^{+0.19}_{-0.10}\,\Msun$. This mass measurement is slightly dependent on the fall height of material from the inner accretion disc, as opposed to IP models that rely on the assumption that in-falling material has reached its free-fall velocity. We take this into account by assuming the detected 424.4\,s signal represents the WD spin period, and we set the WD magnetospheric radius to be equal to the co-rotation radius of the inner accretion disc. This provides the fall-height of accreted material, relative to the mass of the WD as described in Section~3.2. Equating the magnetospheric and co-rotation radius is acceptable for persistently accreting IPs such as IGR~J18007, where the WD and inner accretion disc will have reached spin equilibrium \citep{patterson20}. Even if our interpretation of the 424.4\,s signal is incorrect, this corresponds to a large enough co-rotation radius and a high enough fall-height that the assumption has an insignificant effect on our measurement of the WD mass.

As a further check on our model, we consider an alternative and independent estimate of the shock temperature for IPs. The ratio of Fe\,\textsc{xxv} and Fe\,\textsc{xxvi} ionization states depends on the temperature of the emitting plasma, and is therefore an independent measure of the shock temperature of the accretion column \citep{ezuka99, xu16} and can be used to approximate the WD mass \citep{yu18, xu19}. We calculate that ratio by dividing the measured normalizations of the Gaussian lines at 6.97 and 6.70\,keV to be $0.57\pm0.27$, and this ratio is consistent for both the bremsstrahlung continuum model and our PSR model. Following \citep{xu19}, this would suggest a maximum plasma temperature in the range of 10--40\,keV, which falls just below the lower limit of our \textsc{mkcflow} model temperature at 47\,keV. While estimating the WD mass using the Fe line ratio is not particularly precise, it provides an entirely independent check on the value taken from continuum methods like the PSR model.

In recent years, {\em NuSTAR} observations have been effective constraining the masses of a number of mCVs, due to the high-energy coverage and sensitivity of the instrument \citep{hailey16, suleimanov16, suleimanov19, shaw18, shaw20}. Our measured mass of $1.06^{+0.19}_{-0.10}\,\Msun$ for IGR~J18007 is on the higher side of the mass distribution for mCVs, which have an average mass of $\sim0.8,\Msun$ \citep{shaw20}. Still, we trust our WD mass measurement to be robust, considering that we measure a relatively high bremsstrahlung temperature (only 2 of the sources in \cite{shaw20} have a higher temperature), and considering that similar and higher-mass IPs are common \citep[e.g.][]{tomsick16, deMartino20}. Furthermore, our models provide an excellent statistical fit to the broad spectrum of IGR~J18007 from 0.3--79\,keV rather than relying on only the high-energy spectrum as some studies have \citep{suleimanov18, shaw20}.

Although IGR~J18007 is not near the Chandrasekhar limit, it is still above the average mass of WDs in mCVs, and demonstrates that follow-up studies of CVs discovered by {\em INTEGRAL} will likely have high masses \citep[as another example, see][]{tomsick16}. Understanding the mass distribution of accreting magnetic WDs, especially when compared with the average mass of isolated and non-magnetic WDs ($\sim0.5\,\Msun$), is crucial in explaining the formation and evolution of mCVs \citep{zsg11, shaw20}. For example, continuing to discover massive accreting WDs will provide insight into whether or not accretion and subsequent novae will raise or lower a WD's mass over time. Finally, while CVs are worthy of study on their own, the prospect of accreting WDs as progenitors to type Ia supernovae is of great interest to astrophysics more broadly, and compels us to continue to identify and study sources on the high end of the CV mass distribution.

\section{Conclusion}

We confirm the suggested identity of IGR~J18007 as that of an IP based on our analysis of simultaneous {\em XMM} and {\em NuSTAR} observations alongside optical/IR counterpart photometry. We detect two significant periodic signals in the {\em XMM} data, the stronger of which occurs at $424.4\pm0.7$\,s. This signal has a soft spectrum, and we interpret it as the WD spin period. We also measure the WD mass to be $1.06^{+0.19}_{-0.10}\,\Msun$ based on our spectral modeling. We also find that a soft component is required in order to satisfactorily fit the spectrum, which we model with a single-temperature blackbody of $73^{+8}_{-6}$\,eV. This paper adds IGR~J18007 to the population of well-studied IPs, and further follow-up of similar hard X-ray Galactic sources detected with {\em INTEGRAL} will continue to improve our understanding of these sources.

\section*{Acknowledgments}

B.M.C. acknowledges partial support from the National Aeronautics and Space Administration (NASA) through Chandra Award Numbers GO7-18030X and GO8-19030X issued by the Chandra X-ray Observatory Center, which is operated by the Smithsonian Astrophysical Observatory under NASA contract NAS8-03060. J.A.T. acknowledges partial support from NASA under NuSTAR Guest Observer grant 80NSSC21K0064. M.C. acknowledges financial support from the Centre National d’Etudes Spatiales (CNES). J.H. acknowledges support from an appointment to the NASA Postdoctoral Program at the Goddard Space Flight Center, administered by the Universities Space Research Association under contract with NASA. R.K. acknowledges support from the Russian Science Foundation (grant 19-12-00369).
This work made use of data from the NuSTAR mission, a project led by the California Institute of Technology, managed by the Jet Propulsion Laboratory, and funded by the National Aeronautics and Space Administration. This research has made use of the NuSTAR Data Analysis Software (NuSTARDAS) jointly developed by the ASI Science Data Center (ASDC, Italy) and the California Institute of Technology 
(USA).\\

\section*{Data Availability}

Data used in this paper are available through NASA's HEASARC and the {\em XMM-Newton} Science Archive (XSA).

% BIBLIOGRAPHY

\bibliographystyle{mn2e}
\bibliography{./refs2}{}

\label{lastpage}
\bsp

\end{document}